\documentclass[pra,preprint,amsmath,amssymb,showpacs,showkeys]{revtex4}
\usepackage{bm}
\usepackage{graphics}
\usepackage{epsfig}

\begin{document}

\title{Energy barrier for collapse in a pair of tunnel-coupled
condensates with the scattering lengths of opposite signs}
\author{Valery S. Shchesnovich }
\author{Solange B. Cavalcanti}
\affiliation{ Departamento de F\'{\i}sica - Universidade Federal
de Alagoas, Macei\'o AL 57072-970, Brazil }

\begin{abstract}
We predict the existence of an energy barrier for collapse in a
system of two tunnel-coupled repulsive and attractive quasi
two-dimensional condensates trapped in a double-well potential.
The ground state in such a system can have a lower energy than
that of the collapsing state. We find such ground states
numerically and analytically determine  the domain of their
existence, they have a much larger share of atoms confined in the
repulsive condensate. The energy barrier for collapse  is due to
the fact that the energy of the system  increases if atoms tunnel
from the repulsive condensate to the attractive one.

\end{abstract}
\pacs{03.75.Lm, 03.75.Nt}

\keywords{ Bose-Einstein condensates, double-well trap, collapse,
nonlinear Schr\"odinger equations } \maketitle

\section{Introduction}

Bose-Einstein condensates (BEC) of degenerate quantum gases at
zero temperature are  described with a good accuracy by the
mean-field theory based on the nonlinear Schr\"odinger equation
with an external potential, i.e., the well-known Gross-Pitaevskii
(GP) equation \cite{GPE} for the order parameter. The appearance
of the order parameter is a consequence of the macroscopic quantum
coherence of BEC, which was demonstrated experimentally
\cite{intrfexp,intrf2comp} and explained theoretically
\cite{intrfth} with the use of the GP equation (see also the
review Ref.~\cite{BECRev}).

The interplay between the quantum coherence and nonlinearity in
BEC is a reach source of interesting new phenomena. In this paper
we predict such a new phenomenon related to the collapse
instability of an attractive BEC. The collapse in the condensate
with a negative scattering length is extensively studied
experimentally and theoretically (see, for instance, Refs.
\cite{addCollapse,expCollapseBEC,expCollapseBEC2,theorCollapseBEC,CollapseBEC,ComparTheorExpCollapse}
and the references therein). The multi-component (multi-species)
BECs were also shown to suffer from the collapse instability
\cite{multCollapse1,multCollapse2,multCollapse3,multCollapse4}.
Collapse of a repulsive BEC can be induced by a sudden change of
its positive scattering length to a negative value by application
of the magnetic field near the Feshbach resonance
\cite{Fesh,Fesh2D}.  Here we show that  with the use of a coherent
coupling between two BECs with the opposite interactions (one
repulsive and the other attractive) it is possible to modify
significantly the properties of the collapse instability. An
experimental realization is provided by a pair of condensates
trapped in the double-well potential with far separated wells,
when the atomic scattering length changes sign in just one of the
condensates. The coherent coupling is achieved by the quantum
tunnelling through the central barrier of the double-well trap.

There are two general geometric limits for the condensates trapped
in a double-well potential: the one- and two-dimensional cases. In
the one-dimensional case the condensates  assume the cigar shaped
form, while in the two-dimensional case they have the form of a
pancake.  We consider the two-dimensional case and  rely on the
important fact that the quasi two-dimensional Feshbach resonance
is possible \cite{Fesh2D}.  The one-dimensional nonlinear
Schr\"odinger equation describing the cigar shaped condensate has
no collapsing solutions, they appear only when the transverse
degrees of freedom are taken into account.

The stationary states in  a pair of tunnel-coupled BECs bifurcate
from the zero solution. In this limit the two condensates with
opposite interactions can be described by a single effective
equation of the nonlinear Schr\"odinger type with the nonlinearity
being repulsive or attractive depending on the parameters. When
the effective interaction is repulsive  the system is stable,
whereas, under certain condition on the number of atoms, the
attractive effective interaction causes  collapse.

The surprising fact is, however, that the ground state in a pair
of tunnel-coupled repulsive and attractive condensates can have a
lower energy than the stationary state unstable with respect to
collapse. Therefore, for some values of the parameters, there is
an energy barrier for collapse in the system. Recall that in a
single nonlinear Schr\"odinger equation it is, in fact, the ground
state which suffers from the collapse instability
\cite{theorCollapseBEC,CollapseBEC,ComparTheorExpCollapse}.

This work is related to our previous study \cite{Nova} of the
tunnel-coupled condensates in two spatial dimensions, where we
have considered the stationary states of the system but without
analysis of the ground state. Here we focus on the analysis of the
ground state in the system and its energy, whereas the detailed
derivation of the coupled-mode system and some of the auxiliary
analytical results can be found in our previous publications Refs.
\cite{Nova,DW1D,PhysD}.

The paper is organized as follows. Section \ref{secII} contains
analysis of the ground states in the system in the coupled-mode
approximation, i.e. for far separated wells of the double-well
trap. We conclude with the summary of the results in section
\ref{secIII}. The details of the linear stability analysis of the
stationary solutions are relegated to the Appendix.

\section{Stationary states in the coupled-mode approach}
\label{secII}

Experimental realization of the scattering length management in
just one of the two condensates trapped in a double-well potential
necessitates that the wells be far separated. The lower two energy
levels in such a double-well potential are quasi-degenerate and
there is a  large energy gap separating them from the rest of the
energy spectrum. An ideal quantum gas in such a trap would occupy
the degenerate subspace. For the condensate of an interacting gas
there is a localized basis in the degenerate subspace which allows
one to reduce the three-dimensional Gross-Pitaevskii equation to a
system of linearly coupled two-dimensional equations (see also
Ref. \cite{DW1D}). Indeed, due to the localization of the basis
wave functions in the different wells of the double-well trap, the
nonlinear cross-terms describing the atomic interaction between
the two condensates are negligible.

The central barrier in the double-well trap can be created by a
laser modulation of the parabolic potential. Assuming that the
parabolic trap allows for the pancake shaped condensates, i.e.
that the trap  frequencies satisfy $\gamma =
\omega_z/\omega_\perp\gg1$, we can separate the spatial degrees of
freedom, that is the order parameter for each of the two
condensates factorizes: $\Psi(x,y,z,t) =
\Phi_\perp(x,y,t)\psi_\parallel(z)$, with $\psi_\parallel(z)$
being one of the wave functions of the localized basis. By the
above factorization we neglect the nonlinear term in the
Gross-Pitaevskii equation as compared to the kinetic term along
the double-well trap, i.e. we assume
\begin{equation}
\frac{\hbar^2}{2m}\frac{1}{\ell_z^2} \gg |g||\Psi|^2,
\label{Cond}\end{equation}
where $\ell_z$ is an estimate of the condensate length  in the
$z$-direction and $g = 4\pi a_s/m$ is the nonlinear coefficient in
the Gross-Pitaevskii equation \cite{GPE}. The longitudinal kinetic
term is compensated by the external potential, hence $\ell_z =
\sqrt{{\hbar}/(m\omega_z)}$.

The Hamiltonian  of an atom in the double-well trap can be put in
the form \cite{DW1D,Nova}
\begin{equation}
H = \mathcal{E}|R\rangle\langle R| - K(|R\rangle \langle L| +
|L\rangle\langle R|),
 \label{Ham}\end{equation}
where $|L\rangle$ and $|R\rangle$ are the localized  basis,
$\mathcal{E}$ can be identified with the zero-point energy
difference and $K$ with the tunnelling coefficient. Under the
condition (\ref{Cond}) supplemented by the assumption  of far
separated wells of the trap and the quasi degeneracy of the energy
levels, i.e. $K\ll\hbar\omega_z$ and
$\mathcal{E}\ll\hbar\omega_z$, the Gross-Pitaevskii equation
reduces to the coupled-mode system describing evolution of the
condensates in the transverse dimensions:
\begin{subequations}
\label{EQ1}
\begin{eqnarray}
i\partial_T u &=& \left(-\nabla^2 + \vec{\rho}\,{}^2\right)u -
|u|^2u - \kappa v,
\label{EQ1a}\\
i\partial_T v &=& \left(-\nabla^2  + \vec{\rho}\,{}^2\right)v +
(\varepsilon +  a|v|^2)v - \kappa u
\label{EQ1b}\end{eqnarray}
\end{subequations}
(see also Ref. \cite{Nova}). We have used the dimensionless
variables: $T = \omega_\perp t/2$ and  $\vec{\rho} =
{\vec{r}}/{\ell_\perp}$ (the coordinate in the transverse plane),
with $\ell_\perp =\sqrt{{\hbar}/(m\omega_\perp)}$.  Here $\kappa =
2K/(\hbar\omega_\perp)$, $\varepsilon =
2{\mathcal{E}}/(\hbar\omega_\perp)$, and  $a =
a^{(v)}_s/|a^{(u)}_s|$ is the relative scattering coefficient (we
assume that $a^{(u)}_s<0$ and $a^{(v)}_s>0$).  The term $\rho^2$
in the system (\ref{EQ1}) is due to the common transverse
parabolic trap. Note that due to $\gamma\gg1$ the coefficients
$\kappa$ and $\varepsilon$ are not small in general.

The  transverse order parameters for the two condensates can be
expressed as $\Phi_u = (\sqrt{\Delta}/\ell_\perp)u$ and $\Phi_v =
(\sqrt{\Delta}/\ell_\perp)v$, with $\Delta =
(8\pi|a^{(u)}_s|\int\mathrm{d}z |\psi_\parallel|^4)^{-1}$
\cite{Nova}. For the current experimental traps the parameter
$\Delta$ is on the order of $10^2-10^3$. In the numerics below we
use the quantities $N_{u} = \int\mathrm{d}^2\vec{\rho}\,|u|^2$ and
$N_{v} = \int\mathrm{d}^2\vec{\rho}\,|v|^2$ referring to them as
the number of atoms, for short, though the actual number of atoms
is larger by the factor $\Delta$. Finally, the condition
(\ref{Cond}) in the dimensionless form reads:
\begin{equation}
N_u \ll \gamma R_u^2,\quad aN_v \ll \gamma R_v^2,
\label{Cond1}\end{equation}
where $R_u$ and $R_v$ are  the radii of the two condensates
measured in the units of $\ell_\perp$.

The stationary states with the chemical potential $\mu$ are sought
in the form $u = e^{-i\mu T}U(\rho)$ and $v = e^{-i\mu T}V(\rho)$.
In general, there are two types of the real solutions $U$ and $V$,
satisfying the inequalities $UV>0$ (positive) and $UV<0$
(non-positive).  The positive stationary states bifurcate from
zero at $\mu = \mu_{\mathrm{bif}}$, where $\mu_{\mathrm{bif}} = 2
+ \varepsilon/2 -\sqrt{\varepsilon^2/4 +\kappa^2}$, while the
non-positive ones at $\mu_{\mathrm{bif}}^\prime = 2 +
\varepsilon/2 + \sqrt{\varepsilon^2/4 +\kappa^2}$. The bifurcation
points are easily found by analysis of the linear part of the
stationary coupled-mode system which follows from the system
(\ref{EQ1}) for the vanishing amplitude solutions $U$ and $V$,
given as $U = Ae^{-\rho^2/2}$ and $V = Be^{-\rho^2/2}$ with $B =
(2-\mu)A/\kappa $ and $A\to0$. Being interested in the ground
state of the system, we need to consider only the positive
solutions since they bifurcate from the zero solution at a lower
energy. Below the stationary solution (state) means the positive
solution.

The stationary states were found by using the numerical approach
of Ref. \cite{PhysD}, i.e. by iterative solution in the polar
coordinates of the nonlinear eigenvalue problem arising from the
system (\ref{EQ1}) (with $\mu$ being the nonlinear eigenvalue).
Our method allows for the simultaneous computation of the
stationary state and the chemical potential $\mu$, with the same
accuracy. We have used the Fourier pseudo-spectral method with the
spatial grid containing  256 grid points.

We have found that the sufficient condition for stability of the
stationary state of the system (\ref{EQ1}) is given by the
Vakhitov-Kolokolov (VK) criterion: $\partial N/\partial \mu<0$,
with $N = N_u + N_v$ being the total number of atoms. Moreover, if
the effective interaction is repulsive (see below) then the ground
state solution is unconditionally stable. The applicability of the
VK criterion for stability in the case of the two-dimensional
coupled-mode system is partially based on the numerical analysis
of the appropriate eigenvalue problems, as discussed in the
Appendix.

\begin{figure}[thp]
\epsfig{figure=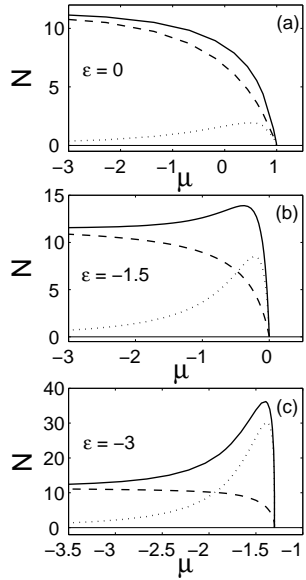,width=0.25\textwidth} \caption{\label{FG1}
The total number of atoms (decreased by the factor $\Delta$) vs.
the chemical potential in the units of $\hbar\omega_\perp/2$ (the
solid line) corresponding to $a = 0.01$ and $\kappa=1$.  The
dashed and dotted lines give the number of atoms in the $u$- and
$v$-condensates, respectively. In panel (a) the unstable
stationary state is attained in the limit of $\mu\to-\infty$,
while in panels (b) and (c) the unstable states appear for finite
negative $\mu$ where $\partial N/\partial\mu>0$. The stable states
correspond to  $\partial N/\partial\mu<0$.}
\end{figure}

If the tunnel-coupling were absent ($\kappa=0$)  the attractive
$u$-condensate would collapse for $N_u
> N_\mathrm{th}\approx 11.69$. This threshold for collapse in the parabolic
trap is given by the number of atoms in the so-called Townes
soliton (see, for instance, Ref. \cite{Collapse2D}). We have found
that in the system (\ref{EQ1}) with $a\ge0$ the stationary states
unstable with respect to collapse always appear in the limit of
large negative $\mu$ and have $N\to N_\mathrm{th}$ as $\mu\to
-\infty$. For some values of the system parameters the collapsing
states exist for finite negative values of $\mu$ (see
Fig.~\ref{FG1}). As $\mu\to -\infty$ they approach the Townes
soliton in the $u$-component, while the $v$-component vanishes
\cite{Nova}. The numerically computed number of atoms as function
of the chemical potential is illustrated  in Fig. \ref{FG1}. For
these values of the parameters $\kappa$ and $a$, the behavior of
the number of atoms for positive values of the zero point energy
difference is similar to that of figure \ref{FG1}(a). It is seen
that the curve $N = N(\mu)$ develops a local maximum close to the
bifurcation point as the zero-point energy difference decreases.
This is related to the existence of a threshold value of the
energy difference, given as $\varepsilon_{0} = {\kappa^2}/{2} -
2$, see below.

Let us determine the energy of the stationary states. In the
coupled-mode approximation the energy (measured in the units of
$\Delta\hbar\omega_\perp/2$) reads
\begin{widetext}
\begin{equation}
E = \int\mathrm{d}^2\vec{\rho}\, \left\{ |\nabla u|^2 + |\nabla
v|^2 + \rho^2(|u|^2 +|v|^2) +\varepsilon |v|^2 - \kappa(uv^* +
vu^*) - \frac{|u|^4}{2} + a\frac{|v|^4}{2} \right\}.
\label{EQ2}\end{equation}
\end{widetext}
Using the coupled-mode system (\ref{EQ1}) to express the kinetic
energy of a stationary state one can reduce the expression for the
total energy  to the following
\begin{equation}
E = \mu N + \frac{1}{2}\int\mathrm{d}^2\vec{\rho}\, \left\{U^4 -
aV^4 \right\}.
\label{EQ3}\end{equation}

\begin{figure}[tbp]
\epsfig{figure=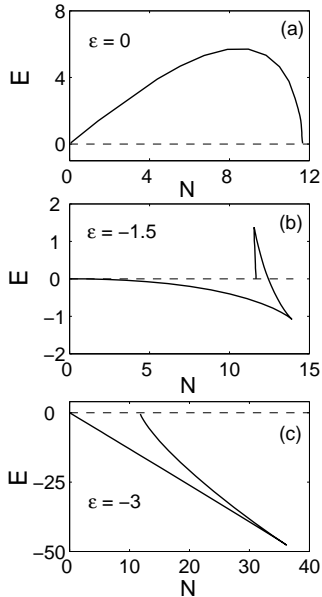,width=0.275\textwidth}
\caption{\label{FG2} The energy of the stationary state (in the
units of $\Delta\hbar\omega_\perp/2$) vs. the total number of
atoms (decreased by the factor $\Delta$) corresponding to figure
1. The almost straight line in panel (c) contains the ground state
of the system stable with respect to collapse. }
\end{figure}

\begin{figure}[th]
\epsfig{figure=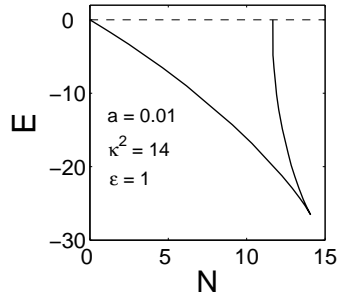,width=0.275\textwidth}
\caption{\label{FG4} An example of the energy barrier for collapse
in the system with the attractive effective interaction. In this
case $\varepsilon_{\mathrm{int}} = -10.56$, while $\varepsilon_{0}
= 5$. Here the axis units are as in figure \ref{FG2}.}
\end{figure}

The  numerically computed energy of the stationary state vs.  the
total number of atoms, corresponding to Fig.~\ref{FG1}, is
illustrated in Fig.~\ref{FG2} (to produce this figure we have
computed the number of atoms and the energy separately by using
the numerically found stationary states). It is seen that the
collapsing states with $N$ approaching the limit value
$N_{\mathrm{th}}$ have the energy approaching zero. Note the
change of sign of the derivative ${\partial E}/{\partial N}$ at
$N=0$ as we decrease the zero-point energy difference in panels
(a) through (c). (The fact that the zero-point energy difference
is negative in figures \ref{FG2}(b) and (c) is a mere coincidence,
see also figure \ref{FG4}.)

The derivative ${\partial E}/{\partial N}$ at $N=0$ can be found
analytically as follows.  For $\epsilon = \mu_{\mathrm{bif}} - \mu
\to0$ the balance of the terms $U^3$ and $\mu U$ and the linear
coupling of the $U$ and $V$ components in equation (\ref{EQ1a})
give $U,V \sim \sqrt{|\epsilon|}$, while the characteristic radius
of the solution is fixed by the trap size. Hence, from equation
(\ref{EQ3}) we conclude that $E = \mu_{\mathrm{bif}} N +
\mathcal{O}(\epsilon^2)$ due to the fact that $N \sim |\epsilon|$
in this limit. Thus  ${\partial E}/{\partial N}=
\mu_{\mathrm{bif}} + \mathcal{O}(\epsilon)$.

If the sign of ${\partial E}/{\partial N}$ at $N=0$  is negative,
that is $\varepsilon\le\varepsilon_{0}$ with $\varepsilon_{0} =
{\kappa^2}/{2} - 2$, and the maximal value $N_\mathrm{m}$ of the
number of atoms in a stationary state exceeds $N_\mathrm{th}$ then
the ground state of the system with $N>N_\mathrm{th}$ is secured
from collapse by an energy barrier.   Indeed, first of all as
either $\mu\to-\infty$ or $\mu\to\mu_\mathrm{bif}$ we have
$E\to0$. Hence, for $\mu_\mathrm{bif} <0$ the function $E = E(N)$
has a local minimum at $N_\mathrm{m}$, where it assumes a negative
value. Consider the stable stationary solutions with negative
energy and the number of atoms $N_\mathrm{th}<N<N_\mathrm{m}$
(such solutions lie on the almost straight line of the curve
$E=E(N)$ in Fig. \ref{FG2}(c)). The system in such a state can
collapse if  some finite amount of energy is passed to the
condensates by an external perturbation, i.e. there is an energy
barrier for collapse.

\begin{figure}[th]
\epsfig{figure=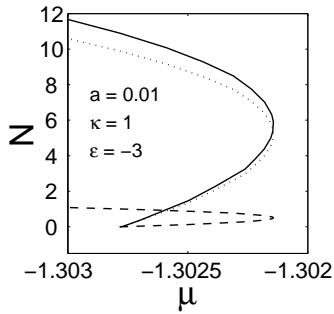,width=0.275\textwidth}
\caption{\label{FG3} Zoomed in part of figure \ref{FG1}(c) about
$\mu = \mu_{\mathrm{bif}}$ which manifests the effective repulsion
in the coupled-mode system  in the vicinity of the bifurcation
point.  Here the axis units are as in figure \ref{FG1}. }
\end{figure}

In an attempt to understand the energy barrier for collapse let us
first try to determine the effective interaction in the
coupled-mode system. The effective interaction depends on the
number of atoms and the double-well trap parameters. An example of
the effective interaction switch from  repulsive to attractive is
given in Fig. \ref{FG3}, which is a  part of Fig.~\ref{FG1}(c)
zoomed in about $\mu_\mathrm{bif}$. The interaction changes sign
at ${\partial \mu}/{\partial N}= 0$ when a zero eigenvalue appears
the linear stability spectrum (see Eqs. (\ref{EQ28})-(\ref{EQ30})
in the Appendix; the positivity of the linear stability spectrum
is a sign of the effective repulsion in the system, see below). As
shown in the Appendix, the stationary states are unconditionally
stable when the effective interaction is repulsive.

The effective interaction can be analytically studied as $\epsilon
= \mu_{\mathrm{bif}}-\mu \to0$. First we use equation (\ref{EQ1b})
to express $V$ in terms of $U$ and its derivatives as follows
\begin{equation}
V = \kappa\beta U - a\kappa^3\beta^4 U^3 +\kappa\beta^2(\nabla^2
+2 - \rho^2) U + 3a^2\kappa^5\beta^7U^5 + \ldots,
\label{EQ3.5}\end{equation}
where $\beta = (\varepsilon + 2 - \mu)^{-1}$ and the higher order
terms are represented by dots. Equation (\ref{EQ3.5}) is derived
in a similar way as an analogous equation for the one-dimensional
coupled mode system in Ref. \cite{DW1D}, one only has to replace
the second-order differential operator by $\nabla^2 +2 - \rho^2$
and $\mu$ by $\mu-2$. Following Ref. \cite{DW1D} we substitute the
expression (\ref{EQ3.5})  into equation (\ref{EQ1a}) and get
\begin{widetext}
\begin{equation}
-\epsilon\left(1 + \frac{\kappa^2}{\mu_0^2}\right)U + \left(1 +
\frac{\kappa^2}{\mu_0^2}\right)(\nabla^2 +2 - \rho^2)U + \left(1 -
\frac{a\kappa^4}{\mu_0^4}\right)U^3 +
\frac{3a^2\kappa^6}{\mu_0^7}U^5 + \ldots  = 0,
\label{EQ4}\end{equation}
\end{widetext}
where $\mu_0 \equiv 2 + \varepsilon/2 +\sqrt{\varepsilon^2/4
+\kappa^2}$.

In the limit of the small amplitude solution the interaction sign
is defined by the coefficient at the cubic term (apart from the
exceptional case of zero). The effective interaction is attractive
for $\varepsilon>\varepsilon_{\mathrm{int}}$ with
\mbox{$\varepsilon_{\mathrm{int}} = \kappa(a^{1/4} - a^{-1/4})$.}
This formula accounts for Fig.~\ref{FG3},  since there we have
$\varepsilon <\varepsilon_{\mathrm{int}} = -2.85$.

Comparison of Figs.~\ref{FG2}(c) and~\ref{FG3} leads to the
conclusion that the energy barrier for collapse in the system
cannot be explained by the repulsive effective interaction.
Indeed, the interval of the $N$ axis occupied by the stable ground
states in Fig. \ref{FG2}(c) is much wider that of the repulsive
interaction in Fig. \ref{FG3}.

Moreover, we get $\varepsilon_{\mathrm{int}}\le \varepsilon_{0}$
for $\kappa \ge \kappa_{\mathrm{cr}}$, where
$\kappa_{\mathrm{cr}}\equiv \sqrt{Q^2 + 4} - Q$ with $Q = a^{-1/4}
- a^{1/4}$.  Hence, for $\kappa \ge \kappa_{\mathrm{cr}}$ and
$\varepsilon_{\mathrm{int}}\le \varepsilon\le \varepsilon_{0}$
there are stable ground states with negative energy in the limit
$\mu\to\mu_{\mathrm{bif}}$, while  the effective interaction is
attractive. An example is provided in Fig. \ref{FG4},  where
$\varepsilon = 1$ while $\varepsilon_{\mathrm{int}} = -10.65$ and
$\varepsilon_0 = 5$.

Fig. \ref{FG4} also is a counterexample to the possible
explanation of stability of the  ground state with a negative
energy by the fact that the repulsive condensate has a lower
zero-point energy ($\varepsilon<0$), i.e. is placed in the lower
well of the double-well trap.

To understand the existence and stability of the stationary states
with negative energy let us examine the expression for the energy
(\ref{EQ3}). As shown above, for small number of atoms, i.e. for
$\mu\to\mu_\mathrm{bif}$, the linear part of the r.h.s. of
equation (\ref{EQ3}) is dominant. But for a large number of atoms
the nonlinear part must dominate. In this case, it is seen that
the energy  decreases with increase of the number of atoms if the
term $aV^4$ is dominant in the integral (it is important to note
that the two terms in the integral on the r.h.s. of formula
(\ref{EQ3}) are due to an interplay of the atomic interactions,
the kinetic energy, the quantum tunneling and the zero-point
energy difference in the system, since we have expressed the
kinetic energy of the condensates in terms of the other energies
to obtain formula (\ref{EQ3})). Indeed, comparing the figures
\ref{FG1} and \ref{FG2} we see that when $\partial E/\partial N<0$
a much larger share of atoms is confined in the repulsive
condensate (figures \ref{FG1}(c) and \ref{FG2}(c)). Therefore,
such a stationary state does not collapse, since the energy would
increase if some of the atoms tunnel from the repulsive condensate
to the attractive one. This causes the energy barrier for
collapse.

The stable ground states with negative energy in the system of
tunnel-coupled repulsive and attractive two-dimensional
condensates  are analogous to the ``unusual solitons'' in the
one-dimensional case \cite{DW1D} which also appear close to the
bifurcation point. The unusual soliton is a bright soliton which
has a much larger share of atoms confined in the  repulsive
condensate.

\section{Conclusion}
\label{secIII}

In conclusion, we have found the ground states in the system of
tunnel-coupled repulsive and attractive condensates, which are
stable with respect to collapse. The coupled-mode approximation
have been used and is justified by the condensates being far
separated in the wells of the double-well trap. We predict the
existence of an energy barrier for collapse in the system, which
is related to the appearance of the stable ground state with a
negative energy. The condensates in such a ground state can
collapse only if they are externally excited with a finite amount
of energy being passed to the system.

The ground state  with  a negative energy is characterized by a
large share of atoms trapped in the repulsive condensate. The
contribution to the energy from the  attractive condensate is
positive, whereas from the repulsive one is negative (the
contribution of each condensate is determined as the result of an
interplay between the kinetic energy, the interaction energy and
the zero-point energy, additionally there is the coupling energy
due to the quantum tunneling). Therefore, the energy increases if
some atoms tunnel from the repulsive to the attractive condensate.
This explains the existence of the energy barrier for collapse in
the system. The energy barrier appears for a wide region of the
parameters space under the condition that the zero-point energy
difference is less than a threshold value. The latter depends on
the tunneling coefficient and can be both positive and negative
(i.e. the condensate with repulsion can be placed in the upper
well of the asymmetric double-well trap).

As is known, the Gross-Pitaevskii equation is sufficient to
describe \textit{stable} stationary states of the Bose-Einstein
condensates at zero temperature,  due to the similarity between
the Bogoliubov-de Gennes equations and the equations describing
evolution of a linear perturbation of the condensate order
parameter \cite{APPL}. Therefore, at zero temperature, given that
the cloud of non-condensed atoms can be neglected, the energy
barrier for collapse phenomenon can be observed.

\section{Acknowledgements}

This work was supported by the CNPq-FAPEAL grant of Brazil.

\appendix
\section{Linear stability analysis}

We will consider only the axially symmetric stationary points of
system (\ref{EQ1}): $u = e^{-i\mu T}U(\rho)$ and $v = e^{-i\mu
T}V(\rho)$. Stability can be established by considering the
eigenvalue problem associated with the linearized system. Writing
the perturbed solution as follows
\begin{equation}
u = e^{-i\mu T}\left\{U(\rho) + e^{-i\Omega
T}\mathcal{U}(\vec{\rho})\right\},\quad v = e^{-i\mu
T}\left\{V(\rho) + e^{-i\Omega T}\mathcal{V}(\vec{\rho})\right\},
 \label{EQ27}\end{equation}
where $(\mathcal{U}$,$\mathcal{V})$  is a small perturbation mode
with frequency $\Omega$, one arrives at the following linear
problem for the eigenfrequency:
\begin{equation}
-i\Omega\left(\begin{array}{c}\mathcal{U}_R\\\mathcal{V}_R\end{array}\right)
=\Lambda_0\left(\begin{array}{c}\mathcal{U}_I\\\mathcal{V}_I\end{array}\right),\quad
i\Omega\left(\begin{array}{c}\mathcal{U}_I\\\mathcal{V}_I\end{array}\right)
=\Lambda_1\left(\begin{array}{c}\mathcal{U}_R\\\mathcal{V}_R\end{array}\right),
 \label{EQ28}\end{equation}
with
\begin{equation}
\Lambda_0 = \left(\begin{array}{cc} L^{(u)}_0 & -\kappa \\
-\kappa & L^{(v)}_0\end{array}\right), \quad
\Lambda_1 = \left(\begin{array}{cc} L^{(u)}_1 & -\kappa \\
-\kappa & L^{(v)}_1\end{array}\right).
 \label{EQ29}\end{equation}
Here the scalar operators are defined as follows
\[
L^{(u)}_0 = -(\nabla^2 + U^2 + \mu) + \rho^2,\quad L^{(u)}_1 =
-(\nabla^2 + 3U^2 + \mu) + \rho^2,\quad L^{(v)}_0 = -(\nabla^2 -
aV^2 + \mu - \varepsilon) + \rho^2,
\]
\begin{equation}
\quad L^{(v)}_1 = -(\nabla^2  - 3aV^2 + \mu - \varepsilon) +
\rho^2.
 \label{EQ30}\end{equation}

First of all, the matrix operator $\Lambda_0$ is non-negative for
positive stationary solutions, i.e. satisfying $UV>0$. Indeed, the
scalar operators on the main diagonal of $\Lambda_0$ can be cast
as follows
\[
L^{(u)}_0 = -\frac{1}{U}\nabla U^2\nabla\frac{1}{U} +
\kappa\frac{V}{U},\quad L^{(v)}_0 = -\frac{1}{V}\nabla
V^2\nabla\frac{1}{V} + \kappa\frac{U}{V},
\]
what can be easily verified by direct calculation. Therefore the
scalar product of $\Lambda_0$ with any vector $X =
(X_1(\vec{\rho}),X_2(\vec{\rho}))$  is non-negative:
\[
\langle X|\Lambda_0|X\rangle = \int\mathrm{d}^2\vec{\rho}\,
\left\{X_1^*L^{(u)}_0X_1 + X_2^*L^{(v)}_0X_2 -
\kappa(X_1^*X_2+X_2^*X_1)\right\}
\]
\[
\ge \kappa\int\mathrm{d}^2\vec{\rho}\,\left|X_1\sqrt{\frac{V}{U}}
- X_2\sqrt{\frac{U}{V}}\,\right|^2 \ge 0.
\]
Here we have used the positivity  of the operators $
-\frac{1}{U}\nabla U^2\nabla\frac{1}{U} $ and $ -\frac{1}{V}\nabla
V^2\nabla\frac{1}{V}$. The operator $\Lambda_0$ has one zero mode
given by the stationary point itself: $Z = (U,V)$, $\Lambda_0Z=0$.
Non-negativity of $\Lambda_0$ is an essential fact for the
following. Thus the non-positive solutions, i.e. satisfying the
inequality $UV<0$, are discarded from the consideration.

The lowest eigenfrequency of the linear stability problem can be
found also  by minimizing the following quotient
\begin{equation}
\Omega^2 = \mathrm{min}\frac{\langle X|\Lambda_1|X\rangle}{\langle
X|\Lambda^{-1}_0|X\rangle}
 \label{EQ31}\end{equation}
in the space orthogonal to the zero mode of $\Lambda_0$: $\langle
Z|X\rangle = 0$ (here $\langle X|Y\rangle \equiv
\int\mathrm{d}^2\vec{\rho}\,(X^*_1Y_1 + Y_2X^*_2)$). Equation
(\ref{EQ31}) follows from the eigenvalue problem rewritten as
$\Lambda_0\Lambda_1X = \Omega^2X$ with $X =
(\mathcal{U}_R,\mathcal{V}_R)$.

The imaginary eigenfrequencies $\Omega$, which mean instability,
appear  due to negative eigenvalues of the operator $\Lambda_1$.
If there is just one negative eigenvalue, then the
Vakhitov-Kolokolov (VK) stability criterion ${\partial
N}/{\partial \mu}<0$ applies, which can be established by a simple
repetition of the arguments presented, for instance,  in Ref.
\cite{PhysD}. The limit on the number of negative eigenvalues is
related to the fact that the minimization of the quotient in
equation (\ref{EQ31}) is subject to only one orthogonality
condition, thus only one negative direction  can be eliminated by
satisfying this orthogonality condition.

To apply the  VK  criterion for stability of the stationary points
in the 2D case (as compared to the 1D case \cite{PhysD}) one has
to rely on numerics to establish the number of negative
eigenvalues of the operator $\Lambda_1$. The following simple
strategy was used. The eigenvalue problem was reformulated in the
polar coordinates $(\rho,\theta)$ and the operators were expanded
in Fourier series with respect to $\theta$  by the substitution:
$\nabla^2 \to \nabla^2_\rho - n^2/\rho^2$, where $\nabla^2_\rho =
\partial^2_\rho +\rho^{-1}\partial_\rho$.
Noticing that the orbital operators $\Lambda_{1,n} =
\Lambda_1(\nabla^2 \to \nabla^2_\rho - n^2/\rho^2)$ are ordered as
follows $\Lambda_{1,n+1} \ge \Lambda_{1,n}$, we have checked for
the negative eigenvalues of the first two orbital operators with
$n = 0,1$. It turns out that $\Lambda_{1,1}$ is always positive,
while the operator $\Lambda_{1,0}$ has one negative eigenvalue or
none (the latter corresponds to the effective repulsive
interaction in the system). Thus the VK criterion applies in the
former case, while in the latter one the stationary point is
unconditionally stable.

\end{document}